\documentclass[prl, twocolumn, amsmath, amssymb, showpacs]{revtex4-1}
\usepackage{times}
\usepackage{verbatim}
\usepackage{graphicx}
\usepackage{graphics}
\usepackage{amsmath}
\usepackage{amssymb}
\usepackage[colorlinks,linkcolor=blue,
            citecolor=blue,
            hyperindex,
            pdfstartview=FitH,
            plainpages=false]
            {hyperref}

\begin{document}

\preprint{ }

\title{Electron-phonon Coupling on the Surface of the Topological Insulator Bi$_2$Se$_3$:\\ Determined from Surface Phonon Dispersion Measurements}

\author{Xuetao Zhu$^{1}$, L. Santos$^{2}$, C. Howard$^1$, R. Sankar$^3$, F.C. Chou$^3$, C. Chamon$^1$, M. El-Batanouny$^{1\ast}$}

\affiliation{\sl{$^{1}$Department of Physics, Boston University,
Boston, MA 02215, USA\\$^2$Department of Physics, Harvard
University, Cambridge, MA 02138, USA\\$^3$ Center of Condensed
Matter Sciences, National Taiwan University, Taipei 10617, Taiwan}}
\date{\today}

\begin{abstract}

In this letter we report measurements of the coupling between Dirac
fermion quasiparticles (DFQs) and phonons on the (001) surface of
the strong topological insulator Bi$_2$Se$_3$. While most
contemporary investigations of this coupling have involved examining
the temperature dependence of the DFQ self-energy via angle-resolved
photoemission spectroscopy (ARPES) measurements, we employ inelastic
helium atom scattering to explore, for the first time, this coupling from the phonon
perspective. Using a Hilbert transform, we are able to obtain the
imaginary part of the phonon self-energy associated with a
dispersive surface phonon branch identified in our previous
work \cite{zhu} as having strong interactions with the DFQs. From
this imaginary part of the self-energy we obtain a branch-specific
electron-phonon coupling constant of 0.43, which is stronger than
the values reported form the ARPES measurements.

\end{abstract}

\pacs{63.20.D-, 63.20.K-, 68.49.Bc, 72.10.Di}

\maketitle

Topological insulators (TIs), a recently discovered class of
materials with insulating bulk but exotic Dirac fermion metallic
surface states \cite{Hasan,Qi,Moore1,Fu1,Hasan2}, have become the
focus of intense research by the condensed matter physics community.
The strong spin-orbit coupling in TIs leads to a definite helicity
whereby the spin is locked normal to the wavevector of the surface
electronic state. A fundamental manifest feature of such
spin-textured surface states is their robustness against
spin-independent scattering, an attribute that protects them from
backscattering and localization \cite{Roushan,Zhang3}. In this
sense, the surface states should be very stable in TIs. Indeed,
ARPES and scanning tunneling microscopy (STM) have confirmed that
the topological surface states are protected even against strong
perturbations, provided they are spin-independent \cite{Wray}, even
up to room temperature \cite{Hsieh4}. Consequently, electron-phonon
(e-p) interaction should be the dominant scattering mechanism for
surface Dirac fermions at finite temperatures. Hence, the study of
e-p coupling in TIs is of exceptional importance in assessing
potential applications such as spintronics.

In condensed matter physics the e-p interaction plays a dominant
role in a myriad of phenomena ranging from electrical conductivity
to superconductivity. As such, it has spawned an extensive amount of
literature. E-p interaction changes the dispersion and the lifetime
of both the electronic and phonon states in a material. The effect
of the e-p coupling on the dispersion and lifetime of the states is
contained in the complex self-energy $\Sigma$ (electron) and $\Pi$
(phonon). The real part $\Sigma^\prime$ ($\Pi^\prime$) renormalizes
the dispersion, while the imaginary part $\Sigma^{\prime\prime}$
($\Pi^{\prime\prime}$) accounts for the finite lifetime $\tau$ of
the state arising from the interaction. In spectroscopic terms, the
linewidth (full width at half maximum)
\begin{equation}
\Gamma\,=\,\frac{\hbar}{\tau}\,=\,-2\Sigma^{\prime\prime}(\Pi^{\prime\prime}),
\end{equation}
is frequently used. Because the real and imaginary parts of the
self-energy are related by a Hilbert (or Kramers-Kronig)
transformation, it is sufficient to determine either $\Sigma^\prime$
($\Pi^\prime$) or $\Sigma^{\prime\prime}$ ($\Pi^{\prime\prime}$). We
should note that when determining the self-energy associated with an
electronic state one integrates over all phonon states, and
vice-versa. One can then study the e-p coupling and its consequences
from the electron or phonon perspective. Here we report on our investigation
of the e-p coupling from measurements of phonon dispersions. To the best of
our knowledge this presents the first attempt at such an approach.

In this report we employ the dimensionless
parameter $\lambda$ to quantify the e-p coupling.
It is defined by \cite{Grimvall}
\begin{equation}
\lambda=2\int_0^{\omega_{max}}\frac{\alpha^2F(\omega)}{\omega}d\omega,
\end{equation}
where $\alpha^2F(\omega)$, the so called \'{E}liashberg e-p coupling
function, is a product of an effective e-p coupling $\alpha^2$
involving phonons of energy $\hbar\omega$ and the phonon density of
states $F(\omega)$. Previous attempts to quantify the e-p coupling
from phonon measurements focused on extracting the e-p matrix
element \cite{Guo} rather than $\lambda$, a topic that we reported
on in our previous paper \cite{zhu}.

Because of recent improvements in energy and momentum resolution of
ARPES, it is now possible to obtain detailed information about the
strength of e-p interaction as a function of the energy $\epsilon_i$
and wavevector ${\bf k}$ of electronic states. Yet, extracting such
information on the e-p interaction from experimental data is not
straightforward; it requires approximations that depend on the
properties of the system studied. By measuring the electron's
self-energy in the vicinity of the Fermi energy ($E_F$) as a
function of temperature, one can fit the ensuing
temperature-dependence of lifetime broadening
$\Gamma_{e-p}(\epsilon_i,{\bf k})$ to the relation \cite{Grimvall}
\begin{equation}\label{ep1}
\Gamma_{e-p}(\epsilon_i,{\bf k};T)\,=\,
2\pi\lambda(\epsilon_i,{\bf k})\,k_BT,
\end{equation}
valid at high temperatures, and extract the value of $\lambda$. Yet,
using ARPES experiments to provide data capable of identifying
contributions of individual phonon modes to the e-p coupling remains
elusive \cite{Hofmann}.

The strong three-dimensional TI Bi$_2$Se$_3$, which is the subject
of this report, was found to have a single Dirac cone with a Dirac
point at the SBZ center \cite{Xia,Zhang2}. Because of this
simplicity, it has been extensively studied both experimentally and
theoretically. ARPES studies have been recently extended to cover
e-p coupling in Bi$_2$Se$_3$ \cite{Park3,Valla,Hofmann2}. However,
the results of the reported studies are contradictory. The ARPES
measurements in Ref.~\onlinecite{Park3} and Ref.~\onlinecite{Valla},
using Eq. (\ref{ep1}), conclude that the e-p coupling in
Bi$_2$Se$_3$ is exceptionally weak, with a value of the
dimensionless e-p coupling parameter $\lambda\sim0.08$
\cite{Valla}. Yet, Ref.~\onlinecite{Hofmann2}, using the same ARPES
approach and Eq. (\ref{ep1}), reports a much stronger coupling with
$\lambda\sim0.25$. All of the aforementioned studies are from the
electron perspective, where the e-p interaction is integrated over all phonon modes. By contrast, effects
of the e-p coupling on specific phonon modes are not commonly
investigated.

From the phonon perspective, it is preferable to express the
\'{E}liashberg coupling function in terms of phonon mode linewidths
and frequencies, and write the e-p coupling parameter
$\lambda$ as \cite{Allen1, Allen2, Allen3}
\begin{align}\label{ep}
\lambda\,&=\,\frac{1}{2\pi
N(E_F)}\,\frac{V_a}{(2\pi)^3}\,\int\;d^3q\;\sum_j\,\frac{\gamma_j({\bf
q},\omega_{j{\bf q}})}{\hbar^2\omega^2_{j{\bf q}}} \notag
\\&=\,\frac{V_a}{(2\pi)^3}\,\int\;d^3q\;\sum_j\,\lambda_j({\bf
q},\omega_{j{\bf q}}),
\end{align}
where $N(E_F)$ is the electronic density of states at $E_F$, $V_a$ is the
primitive cell volume in real space, $\gamma_j({\bf q},\omega_{j{\bf q}})$
is the linewidth of the $j$th mode at wavevector ${\bf q}$ in eV,
$\omega_{j{\bf q}}$ is its frequency and $\lambda_j({\bf q},\omega_{j{\bf q}})$
is the mode-specific e-p coupling. $\lambda$ is then proportional to the
sum of the $\lambda_j({\bf q},\omega_{j{\bf q}})$ of the individual
phonon branches averaged over the whole Brillouin zone.

Phonon line broadenings due to e-p coupling are generally small,
even for superconductors with strong coupling. Such small
broadenings are very difficult to detect in both neutron and helium
scattering experiments mainly because of unavoidable instrument
linewidths of a few meVs. This limits possibilities of extracting
e-p broadening contributions from the measured phonon linewidth.
Moreover, in addition to e-p coupling, there are other inherent
contributions to phonon line broadening, such as phonon-phonon
interaction (anharmonicity), phonon-defect scattering and phonon
anti-crossing with other branches. Yet, if $\Pi^\prime$ can be
determined, then it is straightforward to obtain
$\Pi^{\prime\prime}$ with the aid of a Hilbert transform.

Recently we reported experimental and theoretical results of surface
phonon dispersions on Bi$_2$Se$_3$ (001) \cite{zhu}. The most
prominent features are the absence of the acoustic
Rayleigh branch, and the appearance of a low energy isotropic
convexly dispersive surface optical phonon branch, here on denoted
by $\beta$, with an energy maximum of 7.4 meV. It exhibited a
V-shaped minimum at approximately 2{\bf $k_F$}, reflecting a strong
Kohn anomaly. It is the lowest lying surface phonon
branch within 2$k_F$. Our theoretical analysis attributed this
dispersive profile to the renormalization of the surface phonon
excitations by interactions with surface Dirac fermions. The
contribution of the Dirac fermions to this renormalization was
derived in terms of a Coulomb-type perturbation model within the
random phase approximation (RPA), yielding an effective interaction
strength \footnote{In Ref. \onlinecite{zhu} we used the notation
{\bf $\lambda$} to indicate the effective interaction strength
between phonons and DFQs in the Hamiltonian. This should not be
confused with the present discussion of the e-p coupling parameter
$\lambda$, which, despite the similarity in notation, is a different
quantity.} for the $\beta$ branch of about 3.4 eV/{\AA} \cite{zhu}.
In this letter, we show that because the energy renormalization of
the $\beta$ branch is mainly due to e-p coupling, we can simply use
the Hilbert transform to obtain the e-p coupling parameter
$\lambda_\beta$ for this specific surface phonon branch $\beta$.

\begin{figure}[h!]
\begin{center}
\includegraphics[width=0.48\textwidth]{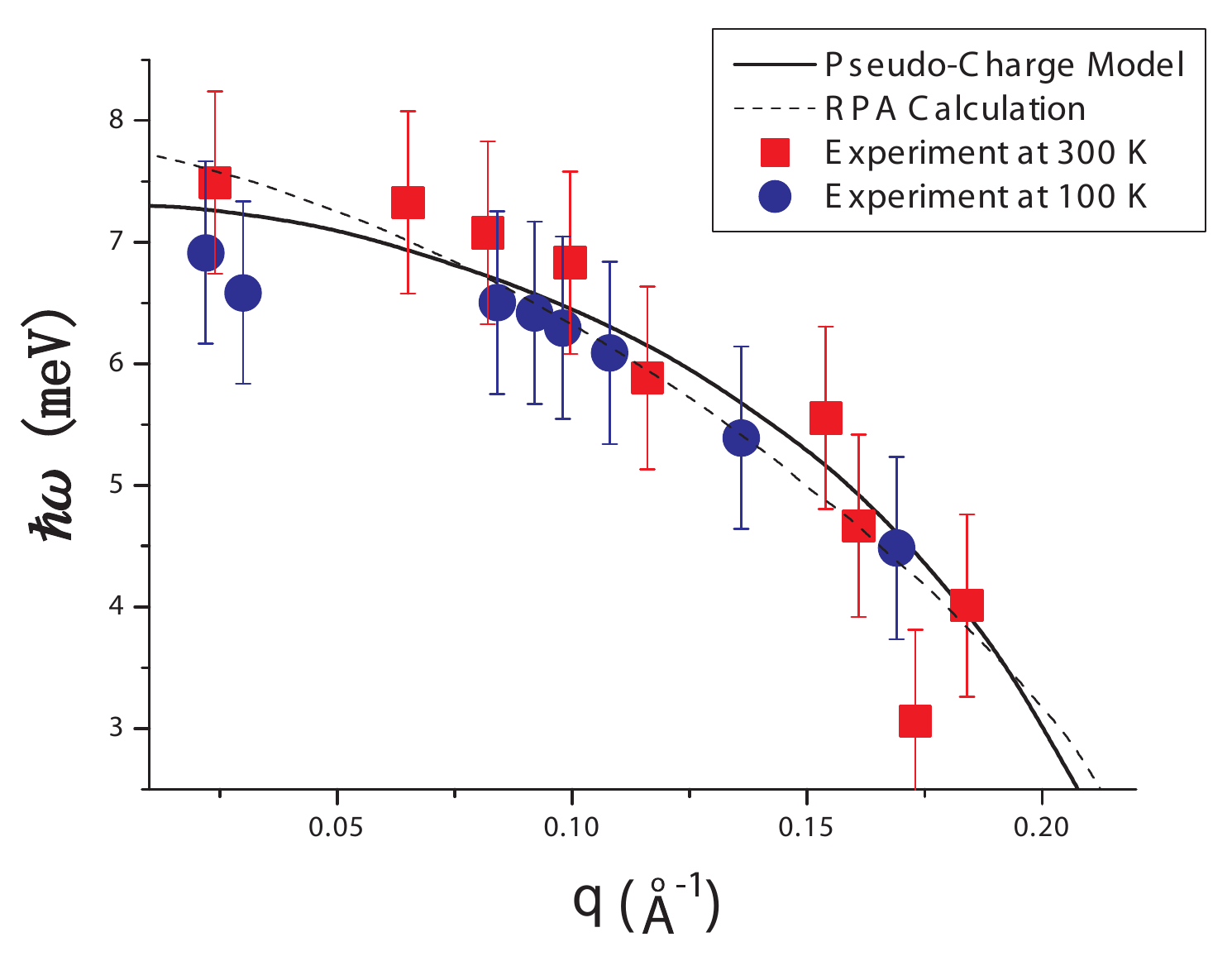}
\caption{\label{fig1} {\small Dispersion curve of the isotropic
phonon branch $\beta$. The red squares represent the experimental
data collected at 300 K, while the blue circles represent the data
collected at 100 K. Vertical error bars indicate the energy
resolution of our facility. The solid line is the result of the
computational implementation of the pseudo-charge model. The dashed
line represents the theoretical RPA calculation of the renormalized
phonon energy \cite{zhu}.}}
\end{center}
\end{figure}

We carried out inelastic helium atom surface scattering (HASS)
measurements for {\it in situ} cleaved (001) surface of
Bi$_2$Se$_3$, at different temperatures in the range 80 K - 300 K.
In Fig. \ref{fig1} we show the experimental data for 300 K and 100 K
(red and blue dots, respectively) for the $\beta$ branch along the
$\overline{\Gamma}$-$\overline{M}$ direction, as well as theoretical
fits obtained with the pseudo-charge model and RPA \cite{zhu}.
Although on average the energy of the 100 K data points is slightly
lower at a given wavevector $q$ than those at 300 K, the difference
is certainly quite small, which justifies carrying out the
theoretical calculations at $T=0$ K. A very good agreement between
theory and experiment is clearly seen.

\begin{figure}[h!]
\begin{center}
\includegraphics[width=0.48\textwidth]{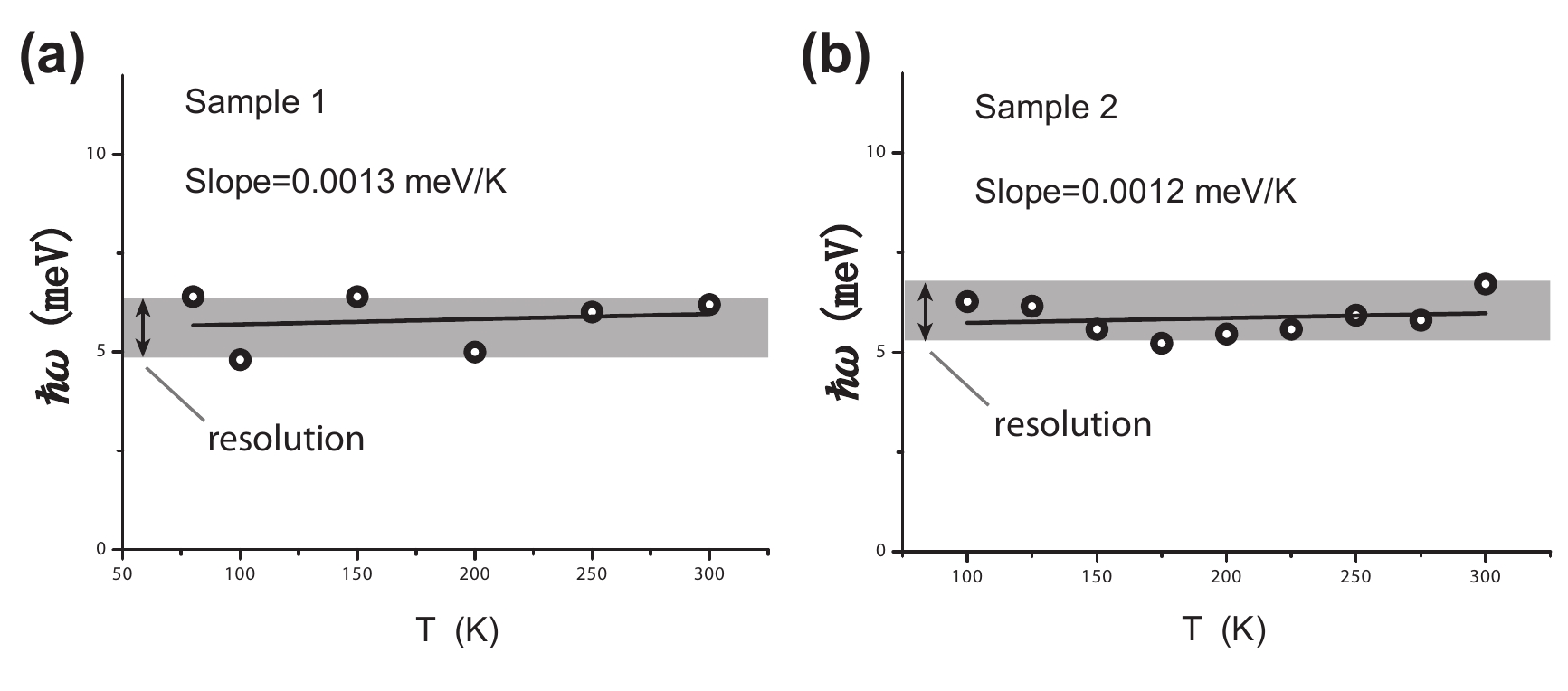}
\caption{\label{E-T} {\small The energy of a single phonon event as
a function of temperature for Sample 1 at $q\sim0.13$ \AA$^{-1}$
{\bf(a)} and for Sample 2 at $q\sim0.6$ \AA$^{-1}$ {\bf(b)},
respectively. The circles represent the experimental data, while the
solid line stands for a linear fit of the data.  The gray band
represents the energy resolution of our facility.}}
\end{center}
\end{figure}

Such measurements were also performed on two different samples at
fixed incident and scattered angles, while varying the temperature.
Fig. \ref{E-T} {\bf (a)} shows the temperature
dependence of a phonon mode of the $\beta$ branch at $q\sim0.13$
\AA$^{-1}$. The phonon energy exhibits a very small increase with
temperature, which is consistent with the trend shown in Fig.
\ref{fig1}. Fig. \ref{E-T} {\bf (b)} shows the temperature
dependence of a phonon mode outside the $\beta$ branch at $q\sim0.6$
\AA$^{-1}$. Consequently, a linear fit to the data has a very small
slope, which justifies neglecting the temperature dependence, in
particular, when we take into account an instrument resolution of
about 1.5 meV. These results demonstrate that
contributions from anharmonicity and defects can be ignored. It
also justifies considering only e-p interaction in our theoretical
RPA calculations.

\begin{figure}[h!]
\begin{center}
\includegraphics[width=0.48\textwidth]{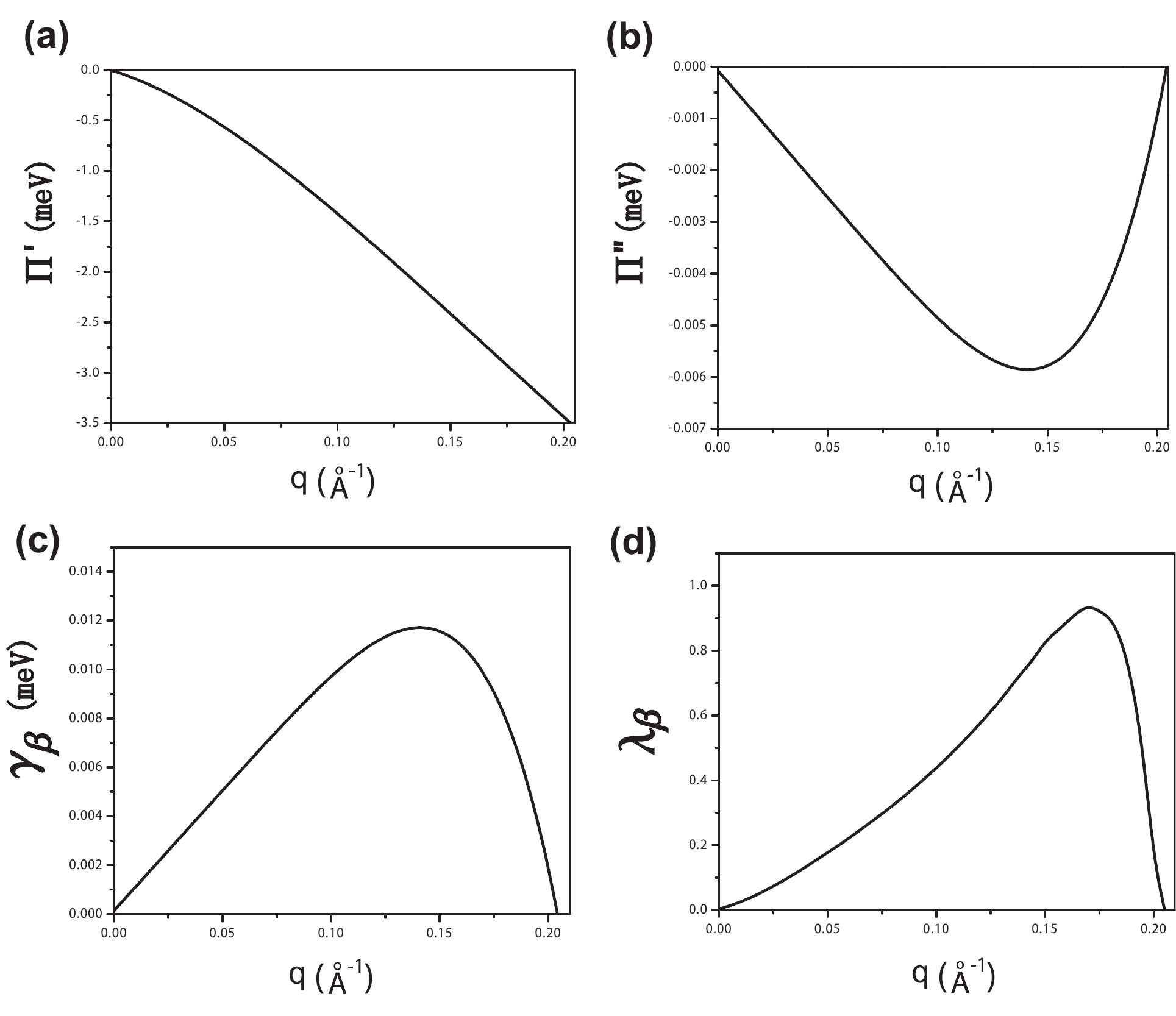}
\caption{\label{ep2} {\small {\bf(a)} The real part of the
self-energy $\Pi^\prime$, and {\bf(b)} its Hilbert transform
$\Pi^{\prime\prime}$, for the isotropic phonon branch $\beta$.
{\bf(c)} The linewidth $\gamma_\beta=-2\Pi^{\prime\prime}$, and
{\bf(d)} the coupling parameter $\lambda_\beta(q)$ determined from
$\gamma_\beta$.}}
\end{center}
\end{figure}

Renormalization of phonon frequencies due to the e-p coupling is
described by the Dyson equation
\begin{equation}\label{dyson}
(\hbar\omega_{{\bf q},\beta})^2 = (\hbar\omega^{(0)}_{{\bf
q},\beta})^2 +2(\hbar\omega^{(0)}_{{\bf q},\beta}) Re \left[
\Pi({\bf q}, \omega_{{\bf q},\beta})\right],
\end{equation}
where $\omega_{{\bf q},\beta}$ and $\omega^{(0)}_{{\bf q},\beta}$
are the renormalized and bare surface phonon frequency for the
$\beta$ branch, respectively, and $\Pi({\bf q}, \omega_{{\bf
q},\beta})$ is the corresponding phonon self-energy. An explicit
expression for $\Pi({\bf q}, \omega_{{\bf q},\beta})$ is given in
Ref.~\onlinecite{zhu} \footnote{In Ref.~\onlinecite{zhu} $\Pi$ is
defined as the the polarization function; it needs to be multiplied
by the prefactor $ \frac{|\text{g}_{{\bf q},\beta}|^2
}{\varepsilon({\bf q}, \omega_{{\bf q},\beta})}$ to become the
self-energy that appears in Eq. (\ref{dyson}).}. We used the
best-fit parameters obtained in Ref.~\onlinecite{zhu} to calculate
$\Pi'({\bf q},\omega_{{\bf q},\beta})$ for the $\beta$ branch at
different wavevectors along the $\overline{\Gamma}$-$\overline{M}$
direction. We obtain the corresponding $\Pi''({\bf q},\omega_{{\bf
q},\beta})$ using the Hilbert transform
\begin{equation}
\Pi''({\bf q},\omega_{{\bf
q},\beta})=\frac{2}{\pi}\int_0^{\infty}\frac{\omega_{{\bf
q},\beta}}{\omega^2_{{\bf q},\beta}-\omega'^2_{{\bf
q},\beta}}\Pi'({\bf q},\omega'_{{\bf q},\beta})d\omega'_{{\bf
q},\beta}.
\end{equation}
We plot the results for $\Pi'$ and $\Pi''$ in panels {\bf (a)} and {\bf (b)} of Fig. \ref{ep2} respectively.

\begin{figure}
\begin{center}
\includegraphics[width=0.4\textwidth]{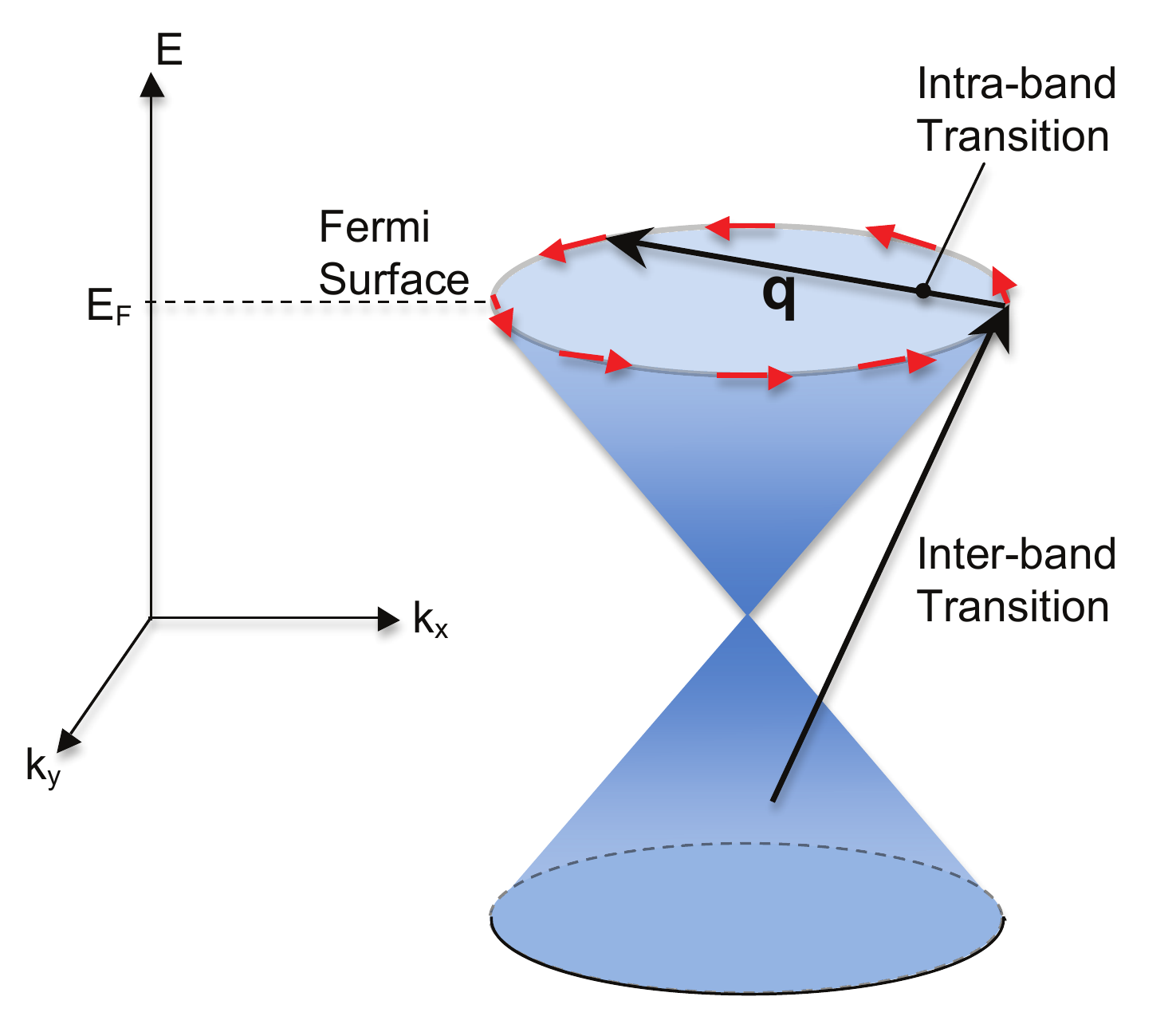}
\caption{\label{DC} {\small Intra- and Inter-band transitions of
DFQs that contribute to the renormalization of the prominent surface
phonon branch $\beta$. ${\bf q}$ is the phonon wave vector. The red
arrows indicate the spin helicity on the Fermi Surface.}}
\end{center}
\end{figure}

The phonon linewidth (FWHM) for the $\beta$ branch, defined as
\cite{Grimvall}
\begin{equation}
\gamma_{\beta}({\bf q})= -2\Pi''({\bf q},\omega_{{\bf q},\beta}),
\end{equation}
is plotted in Fig. \ref{ep2} {\bf (c)} as a function of the
wavevector $q$ along the $\overline{\Gamma}$-$\overline{M}$
direction. We note that the shape of $\gamma_\beta({\bf q})$
reflects the density of the final states for DFQ scattering.
Moreover, the asymmetry apparent in the steep drop as $q$ approaches
$2k_F\sim0.2$ \AA$^{-1}$ stems from the suppression of DFQ
scattering events connecting states with progressively opposing
spins at the far sides of the circular Fermi surface, as shown in
Fig. \ref{DC}.

We obtain the corresponding e-p coupling $\lambda_{\beta}(q)$, shown
in Fig. \ref{ep2} {\bf (d)}, with the aid of the expression
\begin{equation}\label{lambda}
\lambda_{\beta}({\bf q})\,=\,\frac{1}{2\pi
N(E_F)}\,\frac{\gamma_{\beta}({\bf q})}{\hbar^2\omega^2_{{\bf
q},\beta}},
\end{equation}
as derived from Eq. (\ref{ep}). Averaging over the function
$\lambda_{\beta}(q)$, we obtain an effective e-p coupling constant
value for the $\beta$ branch $\lambda_{\beta}=0.43$, which is
greater than the reported values of 0.25 \cite{Hofmann2} and 0.08 \cite{Valla}
obtained from the ARPES measurements. This value of
$\lambda_\beta$, extracted from our experimental data, presents a
lower bound on the actual value of the overall e-p coupling constant
$\lambda$. However, additional contributions from the remaining
higher surface phonon branches to $\lambda$ are not expected to
increase its value significantly above $\lambda_\beta$. The reason
is that the dispersions of each of these higher branches appear the same
regardless of the presence or absence of the DFQs as evidenced from our lattice
dynamics calculations based on the pseudo-charge model shown in figures 2 and 3 in Ref. \cite{zhu}.
Moreover, our result is supported by a recent optical infrared
study \cite{Laforge}, which suggests a strong e-p coupling for the
61 cm$^{-1}$ (or 7.6 meV) optical phonon mode.
Also, a recent theoretical study based on an isotropic elastic continuum phonon model, where e-p coupling involves an acoustic phonon branch, obtains a value of $\lambda \sim0.42$ for thin film geometry \cite {Giraud2}, and $\lambda\sim0.84$ for semi-infinite geometry \cite{Giraud} of Bi$_2$Se$_3$. We note, however, that  our experimental data demonstrates the absence of acoustic surface phonon modes \cite{zhu}.

In summary we use inelastic HASS measurements to determine the e-p
coupling constant for a specific surface phonon branch of
Bi$_2$Se$_3$. Direct experimental measurement of the e-p
contribution to the phonon linewidth is difficult, so instead we
obtain it by extracting the real part of the phonon self-energy
using an RPA fit to the measured surface phonon dispersion curve,
and then obtaining the imaginary part, and hence the e-p
contribution to the phonon linewidth, with the aid of a Hilbert
transform. This approach is supported by the fact that the
experimentally measured temperature independence of the dispersion
demonstrates that the surface phonon frequency renormalization is
mainly determined by e-p coupling. Using this procedure we find an
average e-p coupling parameter for the specific phonon branch
$\lambda_\beta$ = 0.43 that is greater than the values
obtained from the ARPES measurements.

This work is supported by the U.S. Department of Energy
under Grants No. DE-FG02-85ER45222 (MEB) and
DEFG02-06ER46316 (CC). FCC acknowledges the support
from the National Science Council of Taiwan under
project No. NSC 99-2119-M-002-011-MY.

\bibliographystyle{apsrev4-1}
\bibliography{ref}

\begin{thebibliography}{26}%
\makeatletter
\providecommand \@ifxundefined [1]{%
 \@ifx{#1\undefined}
}%
\providecommand \@ifnum [1]{%
 \ifnum #1\expandafter \@firstoftwo
 \else \expandafter \@secondoftwo
 \fi
}%
\providecommand \@ifx [1]{%
 \ifx #1\expandafter \@firstoftwo
 \else \expandafter \@secondoftwo
 \fi
}%
\providecommand \natexlab [1]{#1}%
\providecommand \enquote  [1]{``#1''}%
\providecommand \bibnamefont  [1]{#1}%
\providecommand \bibfnamefont [1]{#1}%
\providecommand \citenamefont [1]{#1}%
\providecommand \href@noop [0]{\@secondoftwo}%
\providecommand \href [0]{\begingroup \@sanitize@url \@href}%
\providecommand \@href[1]{\@@startlink{#1}\@@href}%
\providecommand \@@href[1]{\endgroup#1\@@endlink}%
\providecommand \@sanitize@url [0]{\catcode `\\12\catcode `\$12\catcode
  `\&12\catcode `\#12\catcode `\^12\catcode `\_12\catcode `\%12\relax}%
\providecommand \@@startlink[1]{}%
\providecommand \@@endlink[0]{}%
\providecommand \url  [0]{\begingroup\@sanitize@url \@url }%
\providecommand \@url [1]{\endgroup\@href {#1}{\urlprefix }}%
\providecommand \urlprefix  [0]{URL }%
\providecommand \Eprint [0]{\href }%
\providecommand \doibase [0]{http://dx.doi.org/}%
\providecommand \selectlanguage [0]{\@gobble}%
\providecommand \bibinfo  [0]{\@secondoftwo}%
\providecommand \bibfield  [0]{\@secondoftwo}%
\providecommand \translation [1]{[#1]}%
\providecommand \BibitemOpen [0]{}%
\providecommand \bibitemStop [0]{}%
\providecommand \bibitemNoStop [0]{.\EOS\space}%
\providecommand \EOS [0]{\spacefactor3000\relax}%
\providecommand \BibitemShut  [1]{\csname bibitem#1\endcsname}%
\let\auto@bib@innerbib\@empty
\bibitem [{\citenamefont {Zhu}\ \emph {et~al.}(2011)\citenamefont {Zhu},
  \citenamefont {Santos}, \citenamefont {Sankar}, \citenamefont {Chikara},
  \citenamefont {Howard}, \citenamefont {Chou}, \citenamefont {Chamon},\ and\
  \citenamefont {El-Batanouny}}]{zhu}%
  \BibitemOpen
  \bibfield  {author} {\bibinfo {author} {\bibfnamefont {X.}~\bibnamefont
  {Zhu}}, \bibinfo {author} {\bibfnamefont {L.}~\bibnamefont {Santos}},
  \bibinfo {author} {\bibfnamefont {R.}~\bibnamefont {Sankar}}, \bibinfo
  {author} {\bibfnamefont {S.}~\bibnamefont {Chikara}}, \bibinfo {author}
  {\bibfnamefont {C.~.}\ \bibnamefont {Howard}}, \bibinfo {author}
  {\bibfnamefont {F.~C.}\ \bibnamefont {Chou}}, \bibinfo {author}
  {\bibfnamefont {C.}~\bibnamefont {Chamon}}, \ and\ \bibinfo {author}
  {\bibfnamefont {M.}~\bibnamefont {El-Batanouny}},\ }\href {\doibase
  10.1103/PhysRevLett.107.186102} {\bibfield  {journal} {\bibinfo  {journal}
  {Phys. Rev. Lett.}\ }\textbf {\bibinfo {volume} {107}},\ \bibinfo {pages}
  {186102} (\bibinfo {year} {2011})}\BibitemShut {NoStop}%
\bibitem [{\citenamefont {Hasan}\ and\ \citenamefont {Kane}(2010)}]{Hasan}%
  \BibitemOpen
  \bibfield  {author} {\bibinfo {author} {\bibfnamefont {M.~Z.}\ \bibnamefont
  {Hasan}}\ and\ \bibinfo {author} {\bibfnamefont {C.~L.}\ \bibnamefont
  {Kane}},\ }\href {\doibase 10.1103/RevModPhys.82.3045} {\bibfield  {journal}
  {\bibinfo  {journal} {Rev. Mod. Phys.}\ }\textbf {\bibinfo {volume} {82}},\
  \bibinfo {pages} {3045} (\bibinfo {year} {2010})}\BibitemShut {NoStop}%
\bibitem [{\citenamefont {Qi}\ and\ \citenamefont {Zhang}(2011)}]{Qi}%
  \BibitemOpen
  \bibfield  {author} {\bibinfo {author} {\bibfnamefont {X.-L.}\ \bibnamefont
  {Qi}}\ and\ \bibinfo {author} {\bibfnamefont {S.-C.}\ \bibnamefont {Zhang}},\
  }\href {\doibase 10.1103/RevModPhys.83.1057} {\bibfield  {journal} {\bibinfo
  {journal} {Rev. Mod. Phys.}\ }\textbf {\bibinfo {volume} {83}},\ \bibinfo
  {pages} {1057} (\bibinfo {year} {2011})}\BibitemShut {NoStop}%
\bibitem [{\citenamefont {Moore}(2010)}]{Moore1}%
  \BibitemOpen
  \bibfield  {author} {\bibinfo {author} {\bibfnamefont {J.~E.}\ \bibnamefont
  {Moore}},\ }\href {\doibase 10.1038/nature08916} {\bibfield  {journal}
  {\bibinfo  {journal} {Nature}\ }\textbf {\bibinfo {volume} {464}},\ \bibinfo
  {pages} {194} (\bibinfo {year} {2010})}\BibitemShut {NoStop}%
\bibitem [{\citenamefont {Fu}\ and\ \citenamefont {Kane}(2007)}]{Fu1}%
  \BibitemOpen
  \bibfield  {author} {\bibinfo {author} {\bibfnamefont {L.}~\bibnamefont
  {Fu}}\ and\ \bibinfo {author} {\bibfnamefont {C.~L.}\ \bibnamefont {Kane}},\
  }\href {\doibase 10.1103/PhysRevB.76.045302} {\bibfield  {journal} {\bibinfo
  {journal} {Phys. Rev. B}\ }\textbf {\bibinfo {volume} {76}},\ \bibinfo
  {pages} {045302} (\bibinfo {year} {2007})}\BibitemShut {NoStop}%
\bibitem [{\citenamefont {Hasan}\ and\ \citenamefont {Moore}(2011)}]{Hasan2}%
  \BibitemOpen
  \bibfield  {author} {\bibinfo {author} {\bibfnamefont {M.~Z.}\ \bibnamefont
  {Hasan}}\ and\ \bibinfo {author} {\bibfnamefont {J.~E.}\ \bibnamefont
  {Moore}},\ }\href {\doibase 10.1146/annurev-conmatphys-062910-140432}
  {\bibfield  {journal} {\bibinfo  {journal} {Annual Review of Condensed Matter
  Physics}\ }\textbf {\bibinfo {volume} {2}},\ \bibinfo {pages} {55} (\bibinfo
  {year} {2011})}\BibitemShut {NoStop}%
\bibitem [{\citenamefont {{Roushan}}\ \emph {et~al.}(2009)\citenamefont
  {{Roushan}}, \citenamefont {{Seo}}, \citenamefont {{Parker}}, \citenamefont
  {{Hor}}, \citenamefont {{Hsieh}}, \citenamefont {{Qian}}, \citenamefont
  {{Richardella}}, \citenamefont {{Hasan}}, \citenamefont {{Cava}},\ and\
  \citenamefont {{Yazdani}}}]{Roushan}%
  \BibitemOpen
  \bibfield  {author} {\bibinfo {author} {\bibfnamefont {P.}~\bibnamefont
  {{Roushan}}}, \bibinfo {author} {\bibfnamefont {J.}~\bibnamefont {{Seo}}},
  \bibinfo {author} {\bibfnamefont {C.~V.}\ \bibnamefont {{Parker}}}, \bibinfo
  {author} {\bibfnamefont {Y.~S.}\ \bibnamefont {{Hor}}}, \bibinfo {author}
  {\bibfnamefont {D.}~\bibnamefont {{Hsieh}}}, \bibinfo {author} {\bibfnamefont
  {D.}~\bibnamefont {{Qian}}}, \bibinfo {author} {\bibfnamefont
  {A.}~\bibnamefont {{Richardella}}}, \bibinfo {author} {\bibfnamefont {M.~Z.}\
  \bibnamefont {{Hasan}}}, \bibinfo {author} {\bibfnamefont {R.~J.}\
  \bibnamefont {{Cava}}}, \ and\ \bibinfo {author} {\bibfnamefont
  {A.}~\bibnamefont {{Yazdani}}},\ }\href {\doibase 10.1038/nature08308}
  {\bibfield  {journal} {\bibinfo  {journal} {Nature}\ }\textbf {\bibinfo
  {volume} {460}},\ \bibinfo {pages} {1106} (\bibinfo {year}
  {2009})}\BibitemShut {NoStop}%
\bibitem [{\citenamefont {Zhang}\ \emph
  {et~al.}(2009{\natexlab{a}})\citenamefont {Zhang}, \citenamefont {Cheng},
  \citenamefont {Chen}, \citenamefont {Jia}, \citenamefont {Ma}, \citenamefont
  {He}, \citenamefont {Wang}, \citenamefont {Zhang}, \citenamefont {Dai},
  \citenamefont {Fang}, \citenamefont {Xie},\ and\ \citenamefont
  {Xue}}]{Zhang3}%
  \BibitemOpen
  \bibfield  {author} {\bibinfo {author} {\bibfnamefont {T.}~\bibnamefont
  {Zhang}}, \bibinfo {author} {\bibfnamefont {P.}~\bibnamefont {Cheng}},
  \bibinfo {author} {\bibfnamefont {X.}~\bibnamefont {Chen}}, \bibinfo {author}
  {\bibfnamefont {J.-F.}\ \bibnamefont {Jia}}, \bibinfo {author} {\bibfnamefont
  {X.}~\bibnamefont {Ma}}, \bibinfo {author} {\bibfnamefont {K.}~\bibnamefont
  {He}}, \bibinfo {author} {\bibfnamefont {L.}~\bibnamefont {Wang}}, \bibinfo
  {author} {\bibfnamefont {H.}~\bibnamefont {Zhang}}, \bibinfo {author}
  {\bibfnamefont {X.}~\bibnamefont {Dai}}, \bibinfo {author} {\bibfnamefont
  {Z.}~\bibnamefont {Fang}}, \bibinfo {author} {\bibfnamefont {X.}~\bibnamefont
  {Xie}}, \ and\ \bibinfo {author} {\bibfnamefont {Q.-K.}\ \bibnamefont
  {Xue}},\ }\href {\doibase 10.1103/PhysRevLett.103.266803} {\bibfield
  {journal} {\bibinfo  {journal} {Phys. Rev. Lett.}\ }\textbf {\bibinfo
  {volume} {103}},\ \bibinfo {pages} {266803} (\bibinfo {year}
  {2009}{\natexlab{a}})}\BibitemShut {NoStop}%
\bibitem [{\citenamefont {{Wray}}\ \emph {et~al.}(2011)\citenamefont {{Wray}},
  \citenamefont {{Xu}}, \citenamefont {{Xia}}, \citenamefont {{Hsieh}},
  \citenamefont {{Fedorov}}, \citenamefont {{Hor}}, \citenamefont {{Cava}},
  \citenamefont {{Bansil}}, \citenamefont {{Lin}},\ and\ \citenamefont
  {{Hasan}}}]{Wray}%
  \BibitemOpen
  \bibfield  {author} {\bibinfo {author} {\bibfnamefont {L.~A.}\ \bibnamefont
  {{Wray}}}, \bibinfo {author} {\bibfnamefont {S.-Y.}\ \bibnamefont {{Xu}}},
  \bibinfo {author} {\bibfnamefont {Y.}~\bibnamefont {{Xia}}}, \bibinfo
  {author} {\bibfnamefont {D.}~\bibnamefont {{Hsieh}}}, \bibinfo {author}
  {\bibfnamefont {A.~V.}\ \bibnamefont {{Fedorov}}}, \bibinfo {author}
  {\bibfnamefont {Y.~S.}\ \bibnamefont {{Hor}}}, \bibinfo {author}
  {\bibfnamefont {R.~J.}\ \bibnamefont {{Cava}}}, \bibinfo {author}
  {\bibfnamefont {A.}~\bibnamefont {{Bansil}}}, \bibinfo {author}
  {\bibfnamefont {H.}~\bibnamefont {{Lin}}}, \ and\ \bibinfo {author}
  {\bibfnamefont {M.~Z.}\ \bibnamefont {{Hasan}}},\ }\href {\doibase
  10.1038/nphys1838} {\bibfield  {journal} {\bibinfo  {journal} {Nature
  Physics}\ }\textbf {\bibinfo {volume} {7}},\ \bibinfo {pages} {32} (\bibinfo
  {year} {2011})}\BibitemShut {NoStop}%
\bibitem [{\citenamefont {Hsieh}\ \emph {et~al.}(2009)\citenamefont {Hsieh},
  \citenamefont {Xia}, \citenamefont {Qian}, \citenamefont {Wray},
  \citenamefont {Dil}, \citenamefont {Meier}, \citenamefont {Osterwalder},
  \citenamefont {Patthey}, \citenamefont {Checkelsky}, \citenamefont {Ong},
  \citenamefont {Fedorov}, \citenamefont {Lin}, \citenamefont {Bansil},
  \citenamefont {Grauer}, \citenamefont {Hor}, \citenamefont {Cava},\ and\
  \citenamefont {Hasan}}]{Hsieh4}%
  \BibitemOpen
  \bibfield  {author} {\bibinfo {author} {\bibfnamefont {D.}~\bibnamefont
  {Hsieh}}, \bibinfo {author} {\bibfnamefont {Y.}~\bibnamefont {Xia}}, \bibinfo
  {author} {\bibfnamefont {D.}~\bibnamefont {Qian}}, \bibinfo {author}
  {\bibfnamefont {L.}~\bibnamefont {Wray}}, \bibinfo {author} {\bibfnamefont
  {J.~H.}\ \bibnamefont {Dil}}, \bibinfo {author} {\bibfnamefont
  {F.}~\bibnamefont {Meier}}, \bibinfo {author} {\bibfnamefont
  {J.}~\bibnamefont {Osterwalder}}, \bibinfo {author} {\bibfnamefont
  {L.}~\bibnamefont {Patthey}}, \bibinfo {author} {\bibfnamefont {J.~G.}\
  \bibnamefont {Checkelsky}}, \bibinfo {author} {\bibfnamefont {N.~P.}\
  \bibnamefont {Ong}}, \bibinfo {author} {\bibfnamefont {A.~V.}\ \bibnamefont
  {Fedorov}}, \bibinfo {author} {\bibfnamefont {H.}~\bibnamefont {Lin}},
  \bibinfo {author} {\bibfnamefont {A.}~\bibnamefont {Bansil}}, \bibinfo
  {author} {\bibfnamefont {D.}~\bibnamefont {Grauer}}, \bibinfo {author}
  {\bibfnamefont {Y.~S.}\ \bibnamefont {Hor}}, \bibinfo {author} {\bibfnamefont
  {R.~J.}\ \bibnamefont {Cava}}, \ and\ \bibinfo {author} {\bibfnamefont
  {M.~Z.}\ \bibnamefont {Hasan}},\ }\href
  {http://dx.doi.org/10.1038/nature08234} {\bibfield  {journal} {\bibinfo
  {journal} {Nature}\ }\textbf {\bibinfo {volume} {460}},\ \bibinfo {pages}
  {1101} (\bibinfo {year} {2009})}\BibitemShut {NoStop}%
\bibitem [{\citenamefont {Grimvall}(1981)}]{Grimvall}%
  \BibitemOpen
  \bibfield  {author} {\bibinfo {author} {\bibfnamefont {G.}~\bibnamefont
  {Grimvall}},\ }\href@noop {} {\emph {\bibinfo {title} {The electron-phonon
  interaction in metals}}}\ (\bibinfo  {publisher} {North-Holland Publishing
  Company},\ \bibinfo {year} {1981})\BibitemShut {NoStop}%
\bibitem [{\citenamefont {Qin}\ \emph {et~al.}(2010)\citenamefont {Qin},
  \citenamefont {Shi}, \citenamefont {Cao}, \citenamefont {Wu}, \citenamefont
  {Zhang}, \citenamefont {Plummer}, \citenamefont {Wen}, \citenamefont {Xu},
  \citenamefont {Gu},\ and\ \citenamefont {Guo}}]{Guo}%
  \BibitemOpen
  \bibfield  {author} {\bibinfo {author} {\bibfnamefont {H.}~\bibnamefont
  {Qin}}, \bibinfo {author} {\bibfnamefont {J.}~\bibnamefont {Shi}}, \bibinfo
  {author} {\bibfnamefont {Y.}~\bibnamefont {Cao}}, \bibinfo {author}
  {\bibfnamefont {K.}~\bibnamefont {Wu}}, \bibinfo {author} {\bibfnamefont
  {J.}~\bibnamefont {Zhang}}, \bibinfo {author} {\bibfnamefont {E.~W.}\
  \bibnamefont {Plummer}}, \bibinfo {author} {\bibfnamefont {J.}~\bibnamefont
  {Wen}}, \bibinfo {author} {\bibfnamefont {Z.~J.}\ \bibnamefont {Xu}},
  \bibinfo {author} {\bibfnamefont {G.~D.}\ \bibnamefont {Gu}}, \ and\ \bibinfo
  {author} {\bibfnamefont {J.}~\bibnamefont {Guo}},\ }\href {\doibase
  10.1103/PhysRevLett.105.256402} {\bibfield  {journal} {\bibinfo  {journal}
  {Phys. Rev. Lett.}\ }\textbf {\bibinfo {volume} {105}},\ \bibinfo {pages}
  {256402} (\bibinfo {year} {2010})}\BibitemShut {NoStop}%
\bibitem [{\citenamefont {Hofmann}\ \emph {et~al.}(2009)\citenamefont
  {Hofmann}, \citenamefont {Sklyadneva}, \citenamefont {Rienks},\ and\
  \citenamefont {Chulkov}}]{Hofmann}%
  \BibitemOpen
  \bibfield  {author} {\bibinfo {author} {\bibfnamefont {P.}~\bibnamefont
  {Hofmann}}, \bibinfo {author} {\bibfnamefont {I.~Y.}\ \bibnamefont
  {Sklyadneva}}, \bibinfo {author} {\bibfnamefont {E.~D.~L.}\ \bibnamefont
  {Rienks}}, \ and\ \bibinfo {author} {\bibfnamefont {E.~V.}\ \bibnamefont
  {Chulkov}},\ }\href {http://stacks.iop.org/1367-2630/11/i=12/a=125005}
  {\bibfield  {journal} {\bibinfo  {journal} {New Journal of Physics}\ }\textbf
  {\bibinfo {volume} {11}},\ \bibinfo {pages} {125005} (\bibinfo {year}
  {2009})}\BibitemShut {NoStop}%
\bibitem [{\citenamefont {{Xia}}\ \emph {et~al.}(2009)\citenamefont {{Xia}},
  \citenamefont {{Qian}}, \citenamefont {{Hsieh}}, \citenamefont {{Wray}},
  \citenamefont {{Pal}}, \citenamefont {{Lin}}, \citenamefont {{Bansil}},
  \citenamefont {{Grauer}}, \citenamefont {{Hor}}, \citenamefont {{Cava}},\
  and\ \citenamefont {{Hasan}}}]{Xia}%
  \BibitemOpen
  \bibfield  {author} {\bibinfo {author} {\bibfnamefont {Y.}~\bibnamefont
  {{Xia}}}, \bibinfo {author} {\bibfnamefont {D.}~\bibnamefont {{Qian}}},
  \bibinfo {author} {\bibfnamefont {D.}~\bibnamefont {{Hsieh}}}, \bibinfo
  {author} {\bibfnamefont {L.}~\bibnamefont {{Wray}}}, \bibinfo {author}
  {\bibfnamefont {A.}~\bibnamefont {{Pal}}}, \bibinfo {author} {\bibfnamefont
  {H.}~\bibnamefont {{Lin}}}, \bibinfo {author} {\bibfnamefont
  {A.}~\bibnamefont {{Bansil}}}, \bibinfo {author} {\bibfnamefont
  {D.}~\bibnamefont {{Grauer}}}, \bibinfo {author} {\bibfnamefont {Y.~S.}\
  \bibnamefont {{Hor}}}, \bibinfo {author} {\bibfnamefont {R.~J.}\ \bibnamefont
  {{Cava}}}, \ and\ \bibinfo {author} {\bibfnamefont {M.~Z.}\ \bibnamefont
  {{Hasan}}},\ }\href {\doibase 10.1038/nphys1274} {\bibfield  {journal}
  {\bibinfo  {journal} {Nat. Phys.}\ }\textbf {\bibinfo {volume} {5}},\
  \bibinfo {pages} {398} (\bibinfo {year} {2009})}\BibitemShut {NoStop}%
\bibitem [{\citenamefont {Zhang}\ \emph
  {et~al.}(2009{\natexlab{b}})\citenamefont {Zhang}, \citenamefont {Liu},
  \citenamefont {Qi}, \citenamefont {Dai}, \citenamefont {Fang},\ and\
  \citenamefont {Zhang}}]{Zhang2}%
  \BibitemOpen
  \bibfield  {author} {\bibinfo {author} {\bibfnamefont {H.}~\bibnamefont
  {Zhang}}, \bibinfo {author} {\bibfnamefont {C.-X.}\ \bibnamefont {Liu}},
  \bibinfo {author} {\bibfnamefont {X.-L.}\ \bibnamefont {Qi}}, \bibinfo
  {author} {\bibfnamefont {X.}~\bibnamefont {Dai}}, \bibinfo {author}
  {\bibfnamefont {Z.}~\bibnamefont {Fang}}, \ and\ \bibinfo {author}
  {\bibfnamefont {S.-C.}\ \bibnamefont {Zhang}},\ }\href
  {http://dx.doi.org/10.1038/nphys1270} {\bibfield  {journal} {\bibinfo
  {journal} {Nat. Phys.}\ }\textbf {\bibinfo {volume} {5}},\ \bibinfo {pages}
  {438} (\bibinfo {year} {2009}{\natexlab{b}})}\BibitemShut {NoStop}%
\bibitem [{\citenamefont {Park}\ \emph {et~al.}(2011)\citenamefont {Park},
  \citenamefont {Jung}, \citenamefont {Han}, \citenamefont {Kim}, \citenamefont
  {Kim}, \citenamefont {Song}, \citenamefont {Koh}, \citenamefont {Kimura},
  \citenamefont {Lee}, \citenamefont {Hur}, \citenamefont {Kim}, \citenamefont
  {Cho}, \citenamefont {Kim}, \citenamefont {Kwon}, \citenamefont {Han},\ and\
  \citenamefont {Kim}}]{Park3}%
  \BibitemOpen
  \bibfield  {author} {\bibinfo {author} {\bibfnamefont {S.~R.}\ \bibnamefont
  {Park}}, \bibinfo {author} {\bibfnamefont {W.~S.}\ \bibnamefont {Jung}},
  \bibinfo {author} {\bibfnamefont {G.~R.}\ \bibnamefont {Han}}, \bibinfo
  {author} {\bibfnamefont {Y.~K.}\ \bibnamefont {Kim}}, \bibinfo {author}
  {\bibfnamefont {C.}~\bibnamefont {Kim}}, \bibinfo {author} {\bibfnamefont
  {D.~J.}\ \bibnamefont {Song}}, \bibinfo {author} {\bibfnamefont {Y.~Y.}\
  \bibnamefont {Koh}}, \bibinfo {author} {\bibfnamefont {S.}~\bibnamefont
  {Kimura}}, \bibinfo {author} {\bibfnamefont {K.~D.}\ \bibnamefont {Lee}},
  \bibinfo {author} {\bibfnamefont {N.}~\bibnamefont {Hur}}, \bibinfo {author}
  {\bibfnamefont {J.~Y.}\ \bibnamefont {Kim}}, \bibinfo {author} {\bibfnamefont
  {B.~K.}\ \bibnamefont {Cho}}, \bibinfo {author} {\bibfnamefont {J.~H.}\
  \bibnamefont {Kim}}, \bibinfo {author} {\bibfnamefont {Y.~S.}\ \bibnamefont
  {Kwon}}, \bibinfo {author} {\bibfnamefont {J.~H.}\ \bibnamefont {Han}}, \
  and\ \bibinfo {author} {\bibfnamefont {C.}~\bibnamefont {Kim}},\ }\href
  {http://stacks.iop.org/1367-2630/13/i=1/a=013008} {\bibfield  {journal}
  {\bibinfo  {journal} {New Journal of Physics}\ }\textbf {\bibinfo {volume}
  {13}},\ \bibinfo {pages} {013008} (\bibinfo {year} {2011})}\BibitemShut
  {NoStop}%
\bibitem [{\citenamefont {{Pan}}\ \emph {et~al.}(2011)\citenamefont {{Pan}},
  \citenamefont {{Fedorov}}, \citenamefont {{Gardner}}, \citenamefont {{Lee}},
  \citenamefont {{Chu}},\ and\ \citenamefont {{Valla}}}]{Valla}%
  \BibitemOpen
  \bibfield  {author} {\bibinfo {author} {\bibfnamefont {Z.~.}\ \bibnamefont
  {{Pan}}}, \bibinfo {author} {\bibfnamefont {A.~V.}\ \bibnamefont
  {{Fedorov}}}, \bibinfo {author} {\bibfnamefont {D.}~\bibnamefont
  {{Gardner}}}, \bibinfo {author} {\bibfnamefont {Y.~S.}\ \bibnamefont
  {{Lee}}}, \bibinfo {author} {\bibfnamefont {S.}~\bibnamefont {{Chu}}}, \ and\
  \bibinfo {author} {\bibfnamefont {T.}~\bibnamefont {{Valla}}},\ }\href@noop
  {} {\bibfield  {journal} {\bibinfo  {journal} {ArXiv e-prints}\ } (\bibinfo
  {year} {2011})},\ \Eprint {http://arxiv.org/abs/1109.3638} {arXiv:1109.3638
  [cond-mat.str-el]} \BibitemShut {NoStop}%
\bibitem [{\citenamefont {Hatch}\ \emph {et~al.}(2011)\citenamefont {Hatch},
  \citenamefont {Bianchi}, \citenamefont {Guan}, \citenamefont {Bao},
  \citenamefont {Mi}, \citenamefont {Iversen}, \citenamefont {Nilsson},
  \citenamefont {Hornek\ae{}r},\ and\ \citenamefont {Hofmann}}]{Hofmann2}%
  \BibitemOpen
  \bibfield  {author} {\bibinfo {author} {\bibfnamefont {R.~C.}\ \bibnamefont
  {Hatch}}, \bibinfo {author} {\bibfnamefont {M.}~\bibnamefont {Bianchi}},
  \bibinfo {author} {\bibfnamefont {D.}~\bibnamefont {Guan}}, \bibinfo {author}
  {\bibfnamefont {S.}~\bibnamefont {Bao}}, \bibinfo {author} {\bibfnamefont
  {J.}~\bibnamefont {Mi}}, \bibinfo {author} {\bibfnamefont {B.~B.}\
  \bibnamefont {Iversen}}, \bibinfo {author} {\bibfnamefont {L.}~\bibnamefont
  {Nilsson}}, \bibinfo {author} {\bibfnamefont {L.}~\bibnamefont
  {Hornek\ae{}r}}, \ and\ \bibinfo {author} {\bibfnamefont {P.}~\bibnamefont
  {Hofmann}},\ }\href {\doibase 10.1103/PhysRevB.83.241303} {\bibfield
  {journal} {\bibinfo  {journal} {Phys. Rev. B}\ }\textbf {\bibinfo {volume}
  {83}},\ \bibinfo {pages} {241303} (\bibinfo {year} {2011})}\BibitemShut
  {NoStop}%
\bibitem [{\citenamefont {Allen}(1972)}]{Allen1}%
  \BibitemOpen
  \bibfield  {author} {\bibinfo {author} {\bibfnamefont {P.~B.}\ \bibnamefont
  {Allen}},\ }\href {\doibase 10.1103/PhysRevB.6.2577} {\bibfield  {journal}
  {\bibinfo  {journal} {Phys. Rev. B}\ }\textbf {\bibinfo {volume} {6}},\
  \bibinfo {pages} {2577} (\bibinfo {year} {1972})}\BibitemShut {NoStop}%
\bibitem [{\citenamefont {Butler}\ \emph {et~al.}(1979)\citenamefont {Butler},
  \citenamefont {Pinski},\ and\ \citenamefont {Allen}}]{Allen2}%
  \BibitemOpen
  \bibfield  {author} {\bibinfo {author} {\bibfnamefont {W.~H.}\ \bibnamefont
  {Butler}}, \bibinfo {author} {\bibfnamefont {F.~J.}\ \bibnamefont {Pinski}},
  \ and\ \bibinfo {author} {\bibfnamefont {P.~B.}\ \bibnamefont {Allen}},\
  }\href {\doibase 10.1103/PhysRevB.19.3708} {\bibfield  {journal} {\bibinfo
  {journal} {Phys. Rev. B}\ }\textbf {\bibinfo {volume} {19}},\ \bibinfo
  {pages} {3708} (\bibinfo {year} {1979})}\BibitemShut {NoStop}%
\bibitem [{\citenamefont {Allen}(1980)}]{Allen3}%
  \BibitemOpen
  \bibfield  {author} {\bibinfo {author} {\bibfnamefont {P.}~\bibnamefont
  {Allen}},\ }\href@noop {} {\emph {\bibinfo {title} {Dynamical Properties of
  Solids}}},\ edited by\ \bibinfo {editor} {\bibfnamefont {G.}~\bibnamefont
  {Horton}}\ and\ \bibinfo {editor} {\bibfnamefont {A.}~\bibnamefont
  {Maradudin}},\ Vol.~\bibinfo {volume} {3}\ (\bibinfo  {publisher}
  {North-Holland Publishing Company},\ \bibinfo {year} {1980})\BibitemShut
  {NoStop}%
\bibitem [{Note1()}]{Note1}%
  \BibitemOpen
  \bibinfo {note} {In Ref. \protect \rev@citealp {zhu} we used the notation
  {\protect \bf $\lambda $} to indicate the effective interaction strength
  between phonons and DFQs in the Hamiltonian. This should not be confused with
  the present discussion of the e-p coupling parameter $\lambda $, which,
  despite the similarity in notation, is a different quantity.}\BibitemShut
  {Stop}%
\bibitem [{Note2()}]{Note2}%
  \BibitemOpen
  \bibinfo {note} {In Ref.~\protect \rev@citealp {zhu} $\Pi $ is defined as the
  the polarization function; it needs to be multiplied by the prefactor $
  \protect \frac {|\protect \text {g}_{{\protect \bf q},\beta }|^2
  }{\varepsilon ({\protect \bf q}, \omega _{{\protect \bf q},\beta })}$ to
  become the self-energy that appears in Eq. (\ref {dyson}).}\BibitemShut
  {Stop}%
\bibitem [{\citenamefont {LaForge}\ \emph {et~al.}(2010)\citenamefont
  {LaForge}, \citenamefont {Frenzel}, \citenamefont {Pursley}, \citenamefont
  {Lin}, \citenamefont {Liu}, \citenamefont {Shi},\ and\ \citenamefont
  {Basov}}]{Laforge}%
  \BibitemOpen
  \bibfield  {author} {\bibinfo {author} {\bibfnamefont {A.~D.}\ \bibnamefont
  {LaForge}}, \bibinfo {author} {\bibfnamefont {A.}~\bibnamefont {Frenzel}},
  \bibinfo {author} {\bibfnamefont {B.~C.}\ \bibnamefont {Pursley}}, \bibinfo
  {author} {\bibfnamefont {T.}~\bibnamefont {Lin}}, \bibinfo {author}
  {\bibfnamefont {X.}~\bibnamefont {Liu}}, \bibinfo {author} {\bibfnamefont
  {J.}~\bibnamefont {Shi}}, \ and\ \bibinfo {author} {\bibfnamefont {D.~N.}\
  \bibnamefont {Basov}},\ }\href {\doibase 10.1103/PhysRevB.81.125120}
  {\bibfield  {journal} {\bibinfo  {journal} {Phys. Rev. B}\ }\textbf {\bibinfo
  {volume} {81}},\ \bibinfo {pages} {125120} (\bibinfo {year}
  {2010})}\BibitemShut {NoStop}%
\bibitem [{\citenamefont {{Giraud}}\ \emph {et~al.}(2011)\citenamefont
  {{Giraud}}, \citenamefont {{Kundu}},\ and\ \citenamefont
  {{Egger}}}]{Giraud2}%
  \BibitemOpen
  \bibfield  {author} {\bibinfo {author} {\bibfnamefont {S.}~\bibnamefont
  {{Giraud}}}, \bibinfo {author} {\bibfnamefont {A.}~\bibnamefont {{Kundu}}}, \
  and\ \bibinfo {author} {\bibfnamefont {R.}~\bibnamefont {{Egger}}},\
  }\href@noop {} {\bibfield  {journal} {\bibinfo  {journal} {ArXiv e-prints}\ }
  (\bibinfo {year} {2011})},\ \Eprint {http://arxiv.org/abs/1111.4063}
  {arXiv:1111.4063 [cond-mat.mes-hall]} \BibitemShut {NoStop}%
\bibitem [{\citenamefont {Giraud}\ and\ \citenamefont {Egger}(2011)}]{Giraud}%
  \BibitemOpen
  \bibfield  {author} {\bibinfo {author} {\bibfnamefont {S.}~\bibnamefont
  {Giraud}}\ and\ \bibinfo {author} {\bibfnamefont {R.}~\bibnamefont {Egger}},\
  }\href {\doibase 10.1103/PhysRevB.83.245322} {\bibfield  {journal} {\bibinfo
  {journal} {Phys. Rev. B}\ }\textbf {\bibinfo {volume} {83}},\ \bibinfo
  {pages} {245322} (\bibinfo {year} {2011})}\BibitemShut {NoStop}%
\end{thebibliography}%

\end{document}